\documentclass[twocolumn,showpacs,prl]{revtex4}
\usepackage{amsfonts,amssymb,amsmath,graphicx}

\begin{document}
\title{Eigenfunction entropy and spectral compressibility for critical random matrix ensembles}
\author{E.~Bogomolny and O.~Giraud}
\affiliation{Univ.~Paris-Sud, CNRS, LPTMS, UMR 8626, Orsay, F-91405, France}
\date{\today}
\begin{abstract}
Based on numerical and perturbation series arguments we conjecture that for certain 
critical random matrix models the information dimension of eigenfunctions $D_1$ and the spectral compressibility $\chi$ are related by the simple equation $\chi+D_1/d=1$, where $d$ is system dimensionality.
\end{abstract}

\maketitle

\textit{Introduction.} -- Recently, there has been considerable interest in the investigation of critical random matrix ensembles (CrRME) (see \cite{seligman, evers, kravtsov} and references therein). CrRME are described by $N\times N$ matrices $M_{mn}$ whose characteristic feature is the slow decrease of off-diagonal matrix elements~\cite{levitov}
\begin{equation}
M_{mn}\sim |m-n|^{-1}. 
\label{slow}
\end{equation} 
Such ensembles were introduced to model the Anderson transition of electrons in a disordered potential. The transition occurs between localized and extended states.
Let $\lambda(\alpha)$  and  $\Psi_{m}(\alpha)$ (labeled throughout the paper by Greek letters) denote the eigenvalues and eigenfunctions of $M_{mn}$.  Localization properties of eigenfunctions can be described by a set of multifractal dimensions $D_q$ defined by
\begin{equation}
\Big \langle \sum_{j=1}^N  |\Psi_j(\alpha)|^{2q}\Big \rangle  \underset{N\to\infty}{\sim}  N^{-(q-1)D_q},
\label{D_q}
\end{equation}
where $\langle \ldots \rangle$ is the average over some eigenvalue window and over random realizations of the matrix. For localized states all $D_q$ are equal to $0$, while for states delocalized over the whole $d$-dimensional space $D_q=d$; for CrRME, states are multifractal and $D_q$ are non-trivial functions of $q$. Statistical properties of the eigenvalues of CrRME can be described by the level compressibility $\chi$. It is defined from the limiting behavior of the spectral number variance  
\begin{equation}
\Sigma^{(2)}(L)=\Big \langle n(L)^2\Big \rangle -\Big \langle n(L)\Big \rangle^2 \underset{L\to\infty}{\sim} \chi L,
\label{def_chi}
\end{equation}
where $n(L)$ is the number of eigenvalues in an interval $L$ (the spectrum is unfolded with mean level spacing 1, so that $\langle n(L)\rangle=L$). For the Poisson statistics of independent random variables $\chi=1$, while for standard random matrix ensembles $\chi=0$; typically for CrRME one has $0<\chi<1$.

Multifractal dimensions $D_q$ are related with eigenfunctions, while compressibility $\chi$ is related with eigenvalues. Thus there is no obvious relation between them. Nevertheless in \cite{chalker} it was argued that for  $d$-dimensional systems one should have $2\chi+D_2/d=1$. Later, it was understood that this relation is valid, in general,  only in a weak multifractality regime, i.e.~at first order in the deviation from the usual random matrix limit $D_2=d$, $\chi=0$~\cite{yevtushenko, yevtushenko_2, mirlin}.

In this Letter we argue that, for three different one-dimensional CrRME considered below as well as for certain two- and three-dimensional systems, the following relation holds:
\begin{equation}
\chi+D_1/d=1.
\label{main_exact}
\end{equation} 
Here $D_1$ is the information dimension, corresponding to the mean eigenfunction entropy averaged over the same window as in (\ref{def_chi}) 
\begin{equation}
\label{entropy}
\Big \langle -\sum_{j=1}^N  |\Psi_j(\alpha)|^{2}\ln |\Psi_j(\alpha)|^{2} \Big \rangle  \underset{N\to\infty}{\sim}  D_1 \ln N.
\end{equation}
We are not aware of general analytical arguments in favor of the conjecture (\ref{main_exact}), as the fractal dimensions are not directly accessible for analytical calculations. Nevertheless, a perturbation series approach provides an analytical way to them. There exist two regimes of perturbation series: strong multifractality when $D_q$ is closed to the Poisson value, $D_q\ll 1$, and weak multifractality when $D_q$ is near the random matrix value, 
$d-D_q\ll 1$ \cite{mirlin}. Note that in the two extreme cases of Poisson and usual random matrices, \eqref{main_exact} is trivially verified. For all CrRME with $d=1$ considered below we checked analytically that at first order of the perturbation series the fractal dimensions have the universal form
\begin{equation}
D_q=\left \{\begin{array}{cl}\dfrac{\Gamma(q-1/2)}{\sqrt{\pi}\,\Gamma(q)}(1-\chi)&\mathrm{for}\; 1-\chi\ll 1,\\
1-q\chi&\mathrm{for}\; \chi\ll 1\end{array}\right . 
\label{universal}
\end{equation} 
in a certain range of values of $q$.  
Thus (\ref{main_exact}) is valid at leading order of perturbation series. To check  relation \eqref{main_exact} for intermediate values we performed careful numerical computations of both $D_1$ and $\chi$. The main result is that with available numerical precision no contradiction with our conjecture has been observed.   
%=============================================

\textit{Critical power law random banded matrices}. -- 
The most investigated CrRME is the ensemble of critical power-law banded random matrices (PLBRM)  \cite{seligman, evers, kravtsov}, \cite{yevtushenko, yevtushenko_2, mirlin}. This is the ensemble of $N\times N$   matrices (real symmetric for $\beta=1$ and complex Hermitian for $\beta=2$) whose matrix elements are independent random Gaussian variables with zero mean and variance (depending on a parameter $b$) given by 
$\langle |H_{nn}|^2 \rangle =\beta^{-1}$ and for $m\neq n$
\begin{equation}
\Big \langle |H_{mn}|^2 \Big \rangle =\frac{1}{2}\left [1+\Big (\frac{m-n}{b}\Big )^2 \right ]^{-1}.
\label{cplbrm}
\end{equation}
In both perturbative regimes of large and small $b$, the fractal dimensions and the level compressibility have been calculated at first order \cite{evers, mirlin}, and it is easy to check that in these regimes (\ref{universal}) is fulfilled for this model.  The second order terms at small $b$ for $\chi$ and $D_2$  have been calculated in \cite{yevtushenko_2} and \cite{yevtushenko} respectively but the result for $D_1$ is yet unknown.  

To check the conjecture (\ref{main_exact}) for intermediate values of $b$ we perform numerical calculations of $D_1$ and $\chi$ for a critical PLBRM where to reduce boundary effects the term $m-n$ in (\ref{cplbrm}) is replaced by $N/\pi \sin((m-n)\pi/N)$ \cite{yevtushenko_2}. The fractal dimension $D_1$ is extracted from a fit of the mean entropy (\ref{entropy}) of the 
form $a+D_1\ln N+b/N$. The mean and variance of the entropy are calculated for eigenvectors of PLBRM of size $N=2^n$, $8\leq n\leq 13$. Average is performed over 8192 eigenvectors (namely $2^{n-3}$ eigenvectors with eigenvalues around the band center, and  $2^{16-n}$ realizations of the random matrices). The  number variance (\ref{def_chi}) is calculated on the unfolded spectrum with mean level spacing $\Delta=1$ by taking the average over windows of length $L=2k\Delta$, $1\leq k\leq 32$, centered at integer positions of the energy $E=-32$ to $E=32$, and over $r$ realizations of the random matrices (from $r=32000$ for $N=256$ to $r=500$ for $n=2048$). The level compressibility $\chi(N)$ is then extracted from a quadratic fit $\Sigma^{(2)}(L)=a+\chi(N) L+cL^2$ in the range $L\in[10,32]$.
The large-$N$ asymptotics for $\chi$ is finally obtained by a linear fit of $\chi(N)$ as a function of $1/N$ over the range $2^8\leq N\leq 2^{11}$.

\begin{figure}
\begin{center}
\includegraphics[width=.76\linewidth]{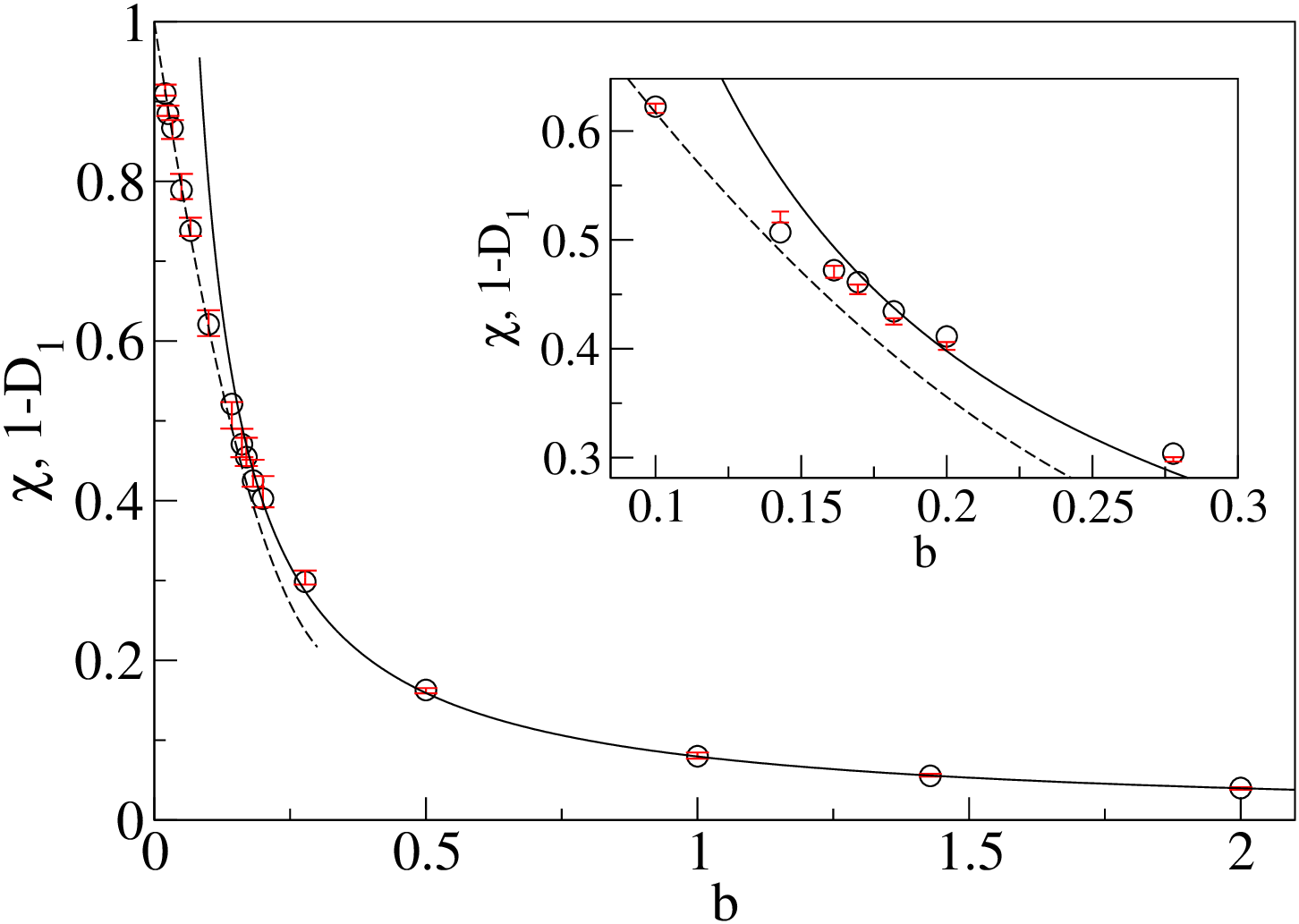}
\end{center}
\caption{(Color online) $\chi$ (black circles) and $1-D_1$ (red error bars) for PLBRM Hermitian ensemble ($\beta=2$). Straight lines: asymptotic theoretical values $\chi=1/(4 \pi b)$ for $b\gg 1$ (solid) and $\chi=1-\pi\sqrt2b-4(2/\sqrt{3}-1)\pi^2 b^2$ for $b\ll 1$ (dashed) ($b^2$ term is taken from \cite{yevtushenko_2} and corrected for a misprint \cite{oleg}). Circles for $\chi$ are larger than error bars. Inset: same data zoomed in.}
\label{fig_plbrm}
\end{figure}

The results for Hermitian matrices ($\beta=2$) are presented in Fig.~\ref{fig_plbrm}. The agreement between $\chi$ and $1-D_1$ is quite good for all $b$. We obtain similar results for the real symmetric case $\beta=1$ (data not shown), which indicates that indeed \eqref{main_exact} holds for PLBRM matrices. 
%=================================================================================

\textit{Ruijsenaars-Schneider ensemble.} -- 
Our second  example of CrRME is the Ruijsenaars-Schneider ensemble (RSE) proposed in \cite{bogomolny}. This ensemble consists of unitary matrices related to the Lax matrix of the Ruijsenaars-Schneider classical $N$-body integrable model. Matrices of this ensemble have the form
\begin{equation}
M_{mn}=\mathrm{e}^{\mathrm{i}\Phi_m}\frac{1-\mathrm{e}^{2\pi\mathrm{ i} a }}{N\Big (1-\mathrm{e}^{2\pi\mathrm{ i}(m-n+a)/N }\Big ) },
\label{RS}
\end{equation}
$1\leq m,n\leq n$. Here $\Phi_m$ are independent random variables (phases) uniformly distributed between $0$ and $2\pi$, $a$ is a free parameter independent on $N$. The advantage of this model is that all spectral correlation functions, in particular the spectral compressibility $\chi$, can be calculated analytically for $N\to\infty$ \cite{bogomolny, bo}. When $0<a<1$, $\chi(a)=(1-a)^2$ \cite{bo}. The general expression for $\chi=\chi(a)$ is tedious and will not be presented here. We only mention that in the vicinity of an integer $k$
\begin{equation}
\chi(a)\underset{a\to k}{\sim}\frac{(a-k)^2}{k^2}.
\label{chi_RS}
\end{equation}

In order to obtain analytical expressions for $D_1$, we construct the perturbation series around all integer points $a=k$. Let $a=k+\varepsilon$ and expand the matrix elements (\ref{RS}) into series of $\varepsilon$. One gets 
\begin{equation}
M_{mn}=M_{mn}^{(0)}\left (1+\frac{\pi \mathrm{i}(N-1)}{N}\varepsilon\right )+\varepsilon M_{mn}^{(1)} +\mathcal{O}(\varepsilon^2)
\label{matrices}
\end{equation}
where $M_{mn}^{(0)}=\mathrm{e}^{\mathrm{i}\Phi_m}\delta_{n,m+k}$,  and
\begin{equation}
M_{mn}^{(1)}=\mathrm{e}^{\mathrm{i}\Phi_m}(1-\delta_{n,m+k})\frac{\pi  \mathrm{e}^{-\pi\mathrm{ i} (m-n+k)/N }}{N\sin(\pi (m-n+k)/N)}
\end{equation}
(here $\delta_{n,m+k}=1$ when $n\equiv m+k$ mod $N$ and 0 otherwise).
In the case $k=0$ (i.e. $|a|\ll 1$), $M_{mn}^{(0)}$ is diagonal: the unperturbed eigenfunctions are localized, and the perturbation decays as in (\ref{slow}). The first order of the perturbation series can be obtained by calculation of contributions from $2\times 2$ sub-matrices \cite{levitov, mirlin, evers}, and one can check that in this case (\ref{universal}) is valid (see \cite{bo} for details). 

\begin{figure}
\begin{center}
\includegraphics[width=.77\linewidth]{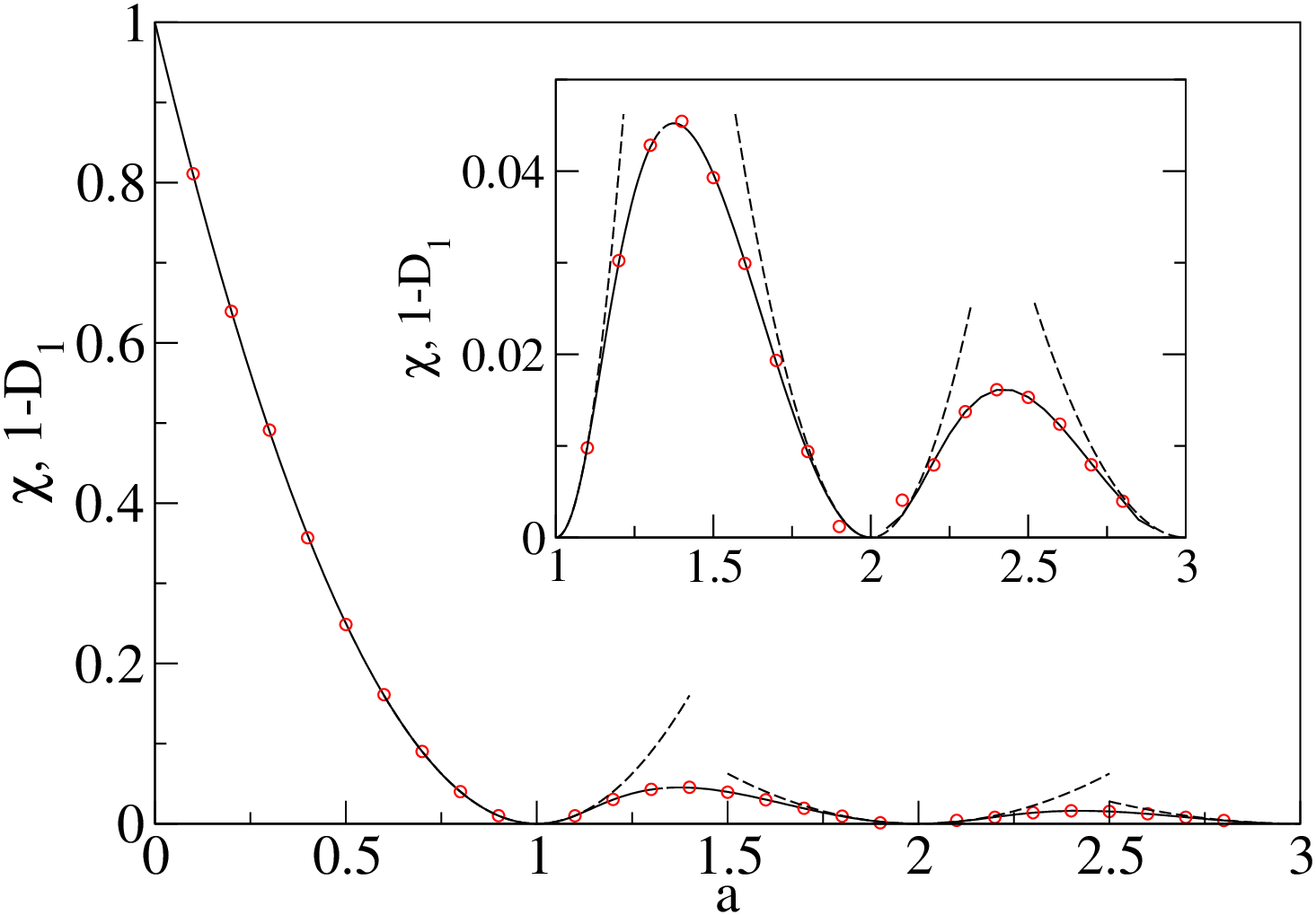}
\end{center}
\caption{(Color online) $\chi$ (solid line) and $1-D_1$ (red circles) for RSE. $D_1$ is obtained numerically by averaging over all eigenvectors taken from $128$ realizations for $N=2^8$ till $8$ for $N=2^{12}$. For $\chi$ we use the theoretical value obtained in \cite{bo}. Dashed line: perturbative regime \eqref{chi_RS}. Inset: zoom for $1\leq a\leq 3$.}
\label{fig_ruij}
\end{figure}

For other values of $k$, eigenfunctions of the unperturbed matrix $M_{mn}^{(0)}$ are extended and the perturbation series corresponds to weak multifractality. Perturbation series in this regime have been constructed for critical PLBRM by using the supermatrix $\sigma$-model \cite{mirlin, evers}. For RSE, the form \eqref{matrices} of the matrix enables us to use standard perturbation series formulas. Let us consider the case $k=1$. Eigenvalues $\lambda^{(0)}(\alpha)$ and eigenfunctions $\Psi_n^{(0)}(\alpha)$ of the unperturbed matrix are given by
\begin{equation}
\lambda^{(0)}(\alpha)=\mathrm{e}^{\mathrm{i}\bar{\Phi}+2\pi \mathrm{i} \alpha/N},\qquad
\Psi_n^{(0)}(\alpha)=\frac{1}{\sqrt{N}}\mathrm{e}^{\mathrm{i}S_n(\alpha)},
\label{theta_alpha}
\end{equation}
where $\bar{\Phi}=\sum_{j=1}^N\Phi_j/N$ and   
\begin{equation}
S_n(\alpha)=\frac{2\pi}{N}\alpha (n-1)-\sum_{j=1}^{n-1}(\Phi_j-\bar{\Phi}).
\label{S_n}
\end{equation}
The expansion of the exact eigenfunctions into a series of unperturbed eigenfunctions has the form
\begin{equation}
\Psi_n(\alpha)=\Psi_n^{(0)}(\alpha)+\sum_{\beta=1}^{N}C_{\alpha \beta}\Psi_n^{(0)}(\beta).
\label{Psi_n_alpha}
\end{equation}
At first order in $\varepsilon=a-1$,
\begin{equation}
C_{\alpha \beta}=\varepsilon\, \frac{\sum_{mn}\Psi_m^{(0)*}(\beta)M_{mn}^{(1)}\Psi_n^{(0)}(\alpha)}{\lambda^{(0)}(\alpha)-\lambda^{(0)}(\beta)}.
\label{C_alpha_beta_1}
\end{equation} 
Expansion at leading order in $\varepsilon$ yields
\begin{equation}
\Big \langle \sum_{n=1}^N |\Psi_n(\alpha)|^{2q}\Big \rangle =N^{1-q}\Big [ 1+\frac{q(q-1)}{2}W(\alpha)\Big ],
\label{second_term}
\end{equation}
where 
\begin{equation}
W(\alpha)=\frac{1}{N}\sum_{n=1}^N \Big \langle \Big [ \sum_{\beta=1}^{N} \mathrm{e}^{\mathrm{i}S_n(\beta)-\mathrm{i}S_n(\alpha)}C_{\alpha \beta}+\mathrm{c.c.}\Big ]^2\Big \rangle . 
\label{W_alpha}
\end{equation}
Using the explicit expressions \eqref{theta_alpha}-\eqref{S_n}
one finds that the exact second order contribution to $W(\alpha)$ is
\begin{equation}
W(\alpha)=
\varepsilon^2\frac{\pi^2}{N^3}\sum_{\beta=1}^{N-1} 
\sum_{n=1}^{N-1}\frac{\sin^2 (\pi \beta n/N)}{\sin^2 ( \pi n/N)\sin^2 (\pi\beta/N)}. 
\label{exact_second}
\end{equation}
We are interested in its behavior for $N\to\infty$. The only diverging terms correspond to two regions. The first is  $\beta \ll N$ with  $n/N$ of the order of 1  and the second is $n\ll N$ and $\beta/N\sim 1$. In this approximation
\begin{equation}
W(\alpha)\underset{N\to \infty}{\sim}
\frac{2\varepsilon^2}{N}\sum_{n=1}^{N-1}\frac{g(n/N)}{\sin^2 ( \pi n/N)},
\end{equation}
where
\begin{equation}
g(n/N)= \sum_{\beta=1}^{\infty}\frac{\sin^2 (\pi \beta n/N)}{\beta^2}=\frac{\pi^2}{2}y(1-y)
\end{equation}     
with $y=n/N$ and $0<y<1$. The remaining sum over $n$  can be transformed into an integral over variable $y$ and finally we obtain
\begin{equation}
W(\alpha)\underset{N\to \infty}{\sim} 2 \varepsilon^2 \ln N+\mathcal{O}(1).
\label{W}
\end{equation}
From (\ref{second_term}) and (\ref{D_q}) it follows that in the leading order of perturbation series in $\varepsilon=1-a$ one has
\begin{equation}
D_q= 1-q(1-a)^2.
\end{equation}
For $k\geq 2$ calculations are more tedious but one can show \cite{bo} that when $|a-k|\ll 1$
\begin{equation}
D_q= 1-q\frac{(a-k)^2}{k^2}.
\end{equation} 
Comparing with (\ref{chi_RS}) we conclude that  the leading terms of perturbation series in RSE indeed verify \eqref{universal} around all integer values of $a$. In Fig.~\ref{fig_ruij} we show that $\chi=1-D_1$ is fulfilled for other values of $a$ as well, with good precision. A variant of RSE has been investigated in \cite{olivier} and the same relation \eqref{main_exact} has been observed to hold within numerical precision. Note that the perturbation series approach used here for weak multifractality can be applied to other problems as well \cite{bo}. 
%===================================================================

\begin{figure}
\begin{center}
\includegraphics[width=.7\linewidth]{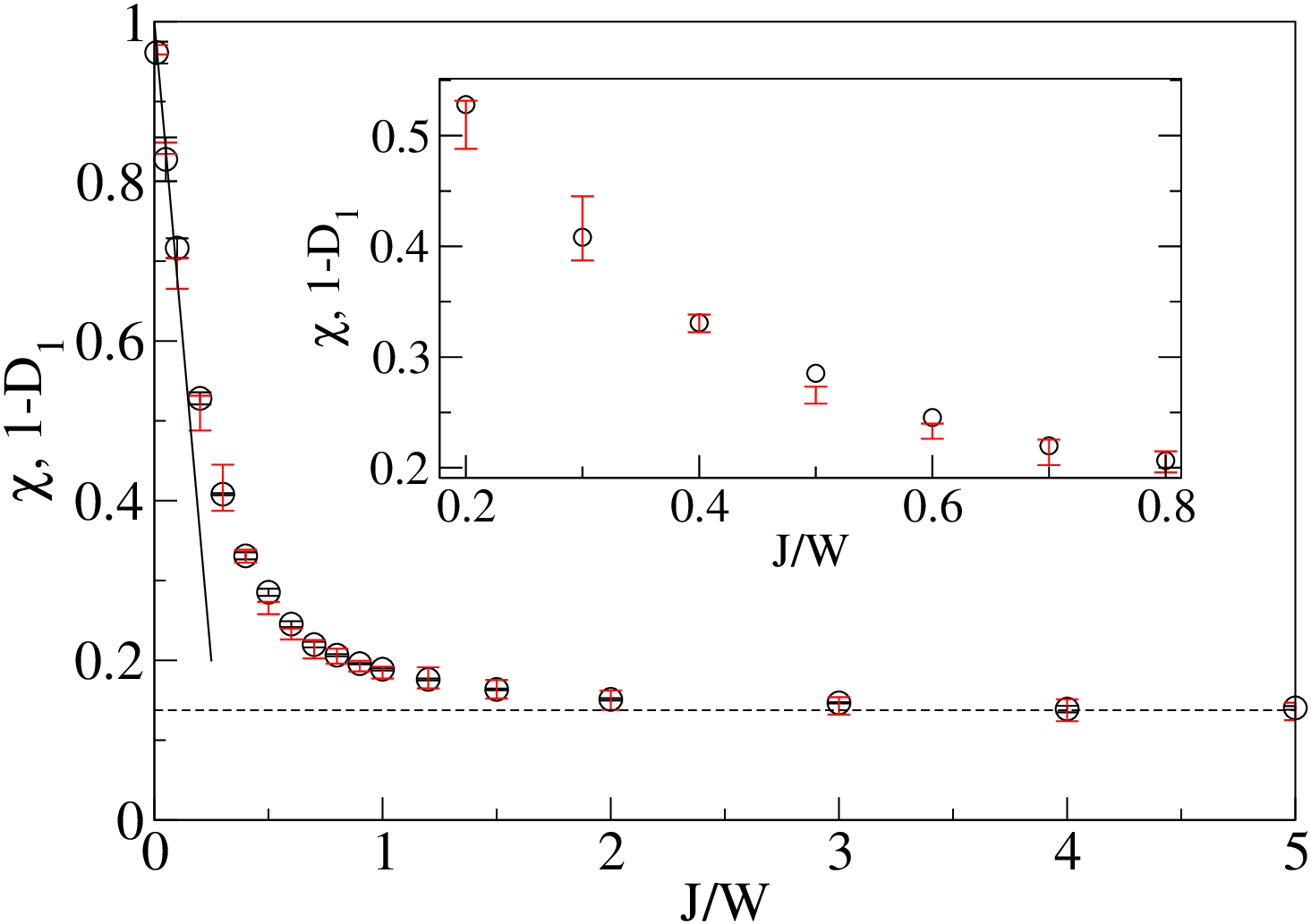}
\end{center}
\caption{(Color online) Same as Fig.~\ref{fig_plbrm} for ultrametric matrices. Solid line is the asymptotic theoretical value \eqref{chi_asympt_ultrametric}. Dashed horizontal line is the numerical value $1-D_1\simeq 0.137\pm 0.008$ obtained for $W=0$ ($J/W\to\infty$). Same method as in Fig.~\ref{fig_plbrm} (here $D_1$ is obtained from matrices of size up to $2^{12}$ only). Inset: same data magnified around $J/W=0.5$. \label{fig_ultrametric}}
\end{figure}

\textit{Critical ultrametric ensemble.} --
As a third example we consider the ensemble of critical ultrametric random matrices proposed in \cite{ossipov}. This ensemble consists of $2^K\times 2^K$  Hermitian matrices whose matrix elements are  independent Gaussian random variables with zero mean. All diagonal elements have the same variance $\langle |H_{nn}|^2 \rangle=W^2$. The variances of off-diagonal elements are $\langle |H_{mn}|^2 \rangle = 2^{2-d_{mn}}J^2$,
where $d_{mn}$ is the ultrametric distance between $m$ and $n$ on the binary tree with $K$ levels and the root at $1$. The parameter in this model is the ratio $J/W$. The first term of perturbation series in $J/W$ for fractal dimensions has been calculated in \cite{ossipov} as 
\begin{equation}
D_q=\frac{J}{W}\frac{\sqrt{\pi}\, \Gamma(q-1/2)}{\sqrt{2}\,\ln 2\,\Gamma(q)}. 
\label{D_q_ultra}
\end{equation}
The calculation of the spectral compressibility can be performed similarly as above \cite{bo} and one finds 
\begin{equation}
\label{chi_asympt_ultrametric}
\chi=1-\frac{\pi\,J}{\sqrt{2} \ln 2\, W}. 
\end{equation} 
Thus the relation $\chi=1-D_1$ is fulfilled for the critical ultrametric ensemble at first order in $J/W$. For other $J/W$, we have calculated $D_1$ and $\chi$ numerically. The results shown in Fig.~\ref{fig_ultrametric} confirm that the relation \eqref{main_exact} is valid with a good accuracy for all values of $J/W$.

%========================================
\textit{Higher dimensional models.} -- 
The above discussion was restricted to one-dimensional systems. Let us now turn to two examples of higher-dimensional systems where numerical data are available in the literature.

The standard two-dimensional critical model  
is the metal-insulator transition in the quantum Hall effect, modeled by the Chalker-Coddington network~\cite{chalker_coddington}. 
We get the information dimension $D_1$ from \cite{evers_mildenberger_mirlin}, where multifractal dimensions for this model have been carefully fitted. Assuming that  errors in the coefficients of the fit are independent we obtain that $D_1=1.7405\pm 0.0004$. According to our conjecture, the compressibility should be $\chi_c=1-D_1/2$, which gives $\chi_c=0.1298\pm 0.0002$. This agrees with the estimate in \cite{klesse_metzler} where for this model the value $\chi =0.124 \pm 0.006$ was obtained numerically. Notice that errors are underestimated as they mostly take into account statistical errors.

The metal-insulator transition in three-dimensional Anderson model is the most important example of critical systems. In \cite{romer} it was reported that $D_1=1.93\pm 0.01$. Assuming that the symmetry for the multifractal spectrum conjectured in \cite{mirlin_fyodorov} holds, then the estimate $D_0^{\prime}=4.027\pm 0.016$ given in \cite{romer_2} yields $D_1=2d-D_0^{\prime}=1.973\pm 0.016$. According to our conjecture (\ref{main_exact}) one should have $\chi_c=1-D_1/3$, which gives $\chi_c\approx 0.34$ to $0.36$. The spectral compressibility for the anisotropic Anderson model at the metal-insulator transition has been reported in \cite{ndawana}. In this paper it was concluded that $\chi=0.28\pm 0.06$ but this value corresponds to the average over different fits which have big fluctuations. In the same paper, when a smaller length has been used to define the number variance it was found that $\chi$ fluctuates much less and gives $\chi=0.32\pm 0.03$, which agrees with $\chi_c\approx 0.35$ obtained from our conjecture. It would be of interest to get $D_1$ and $\chi$ with higher precision for these models.

%=======================================
\textit{Conclusion.} --
In this Letter we present analytical and numerical evidences in favor of the conjecture that for a large class of CrRME the wavefunction entropy and the level compressibility are simply related by $\chi+D_1/d=1$. We consider three different models: the standard critical PLBRM \cite{seligman}, the ensemble of random matrices related with Lax matrices of the Ruijsenaars-Schneider integrable model \cite{bo}, and the critical ultrametric ensemble \cite{ossipov}. For all these models we check the conjecture in perturbation series and by direct numerical calculations. Though we cannot rigorously prove our relation, these investigations show that with good numerical precision it is fulfilled for very different systems. This suggests the existence of an universal structure in CrRME.  

%=======================================
\textit{Acknowledgements} -- EB is greatly indebted to  V.~Kravtsov,  I.~Lerner, A.~Ossipov, and O.~Yevtushenko for useful discussions. He also thanks ICTP for the hospitality during the visit when a part of this work has been done. 
%%%%%%%%%%%%%%%%%%%%%%%%%%%%%%%%%%%%%%%%%%%%%%%%%%%%%%%%%%%%%%%%%%%%%%%%%% 

\end{document}